\documentclass[twocolumn,showpacs,aps,pre]{revtex4}

\usepackage{amssymb,amsmath}
\usepackage[dvips]{graphicx}

\usepackage{hyperref}
\hypersetup{
breaklinks=true,
colorlinks=true,
citecolor=blue,
linkcolor=blue,
filecolor=blue,
urlcolor=blue
}

\begin{document}

\title{Infinite-randomness critical point in the two-dimensional disordered contact process}

\author{Thomas Vojta}
\author{Adam Farquhar}
\author{Jason Mast}
\affiliation{Department of Physics, Missouri University of Science and Technology,
Rolla, MO 65409, USA}

\begin{abstract}
We study the nonequilibrium phase transition in the two-dimensional contact process on a
randomly diluted lattice by means of large-scale Monte-Carlo simulations for times up to
$10^{10}$ and system sizes up to $8000 \times 8000$ sites. Our data provide strong
evidence for the transition being controlled by an exotic infinite-randomness critical
point with activated (exponential) dynamical scaling. We calculate the critical exponents
of the transition and find them to be universal, i.e., independent of disorder strength.
The Griffiths region between the clean and the dirty critical points exhibits power-law
dynamical scaling with continuously varying exponents. We discuss the generality of our
findings and relate them to a broader theory of rare region effects at phase transitions
with quenched disorder. Our results are of importance beyond absorbing state transitions
because according to a strong-disorder renormalization group analysis, our transition
belongs to the universality class of the two-dimensional random transverse-field
Ising model.
\end{abstract}

\date{\today}

\maketitle


\section{Introduction}

Many-particle systems far from equilibrium can undergo phase transitions if external
parameters are varied. These transitions, which separate different nonequilibrium steady
states, are characterized by large-scale fluctuations and collective behavior over large
distances and long times very similar to the behavior at equilibrium critical points.
Examples of nonequilibrium phase transitions occur in a wide variety of systems ranging
from surface chemical reactions and growing interfaces to traffic jams and to the
spreading of epidemics in biology. Reviews of some of these transitions can be found,
e.g., in Refs.\
\cite{ZhdanovKasemo94,SchmittmannZia95,MarroDickman99,Hinrichsen00,Odor04,Luebeck04,TauberHowardVollmayrLee05}

In recent years, a large effort has been directed towards classifying possible
nonequilibrium phase transitions \cite{Odor04}. A particularly well-studied type of
transitions separates active, fluctuating steady states from inactive absorbing states
where fluctuations cease entirely. The generic universality class for these absorbing
state transitions is directed percolation (DP) \cite{GrassbergerdelaTorre79}.
According to a conjecture by Janssen and Grassberger \cite{Janssen81,Grassberger82},
all absorbing state transitions with a scalar order parameter, short-range interactions,
and no extra symmetries or conservation laws belong to this class. Examples include the
transitions in the contact process \cite{HarrisTE74}, catalytic
reactions \cite{ZiffGulariBarshad86}, interface growth \cite{TangLeschhorn92}, or Pomeau's
conjecture regarding turbulence \cite{Pomeau86}.
In the presence of conservation laws or additional symmetries, other universality classes such as
the parity conserving class or the $Z_2$-symmetric directed percolation (DP2) class can occur
(see Ref.\ \cite{Hinrichsen00} and references therein).

The DP universality class is ubiquitous in theory and computer simulations, but clearcut
experimental verifications were lacking for a long time \cite{Hinrichsen00b}. Partial
evidence was manifest in the spatio-temporal intermittency in ferrofluidic spikes
\cite{RuppRichterRehberg03}. Only very recently, a full verification (the only one, to
the best of our knowledge) was found in the transition between two turbulent states
in a liquid crystal \cite{TKCS07}. One possible cause for the surprising rarity of
DP scaling in experiments are impurities, defects, or other kinds of  disorder
that are present in all realistic systems.

For this reason, the effects of quenched spatial disorder on the DP universality class
have been studied for many years. However, a consistent picture has been very slow to
emerge. According to the general Harris criterion \cite{Harris74} (applied to the DP
universality class by Kinzel \cite{Kinzel85} and Noest \cite{Noest86}), a clean critical
point is (perturbatively) stable against weak spatial disorder, if the spatial
correlation length critical exponent $\nu_\perp$ fulfills the inequality $d \nu_\perp
>2$, where $d$ is the spatial dimensionality. The correlation length exponents of the DP
universality class are $\nu_\perp \approx 1.097$ in one space dimension, 0.73 in two
dimensions, and 0.58 in three dimensions \cite{Hinrichsen00}, therefore, the Harris criterion is
violated in all dimensions $d<4$. The instability of the DP critical behavior
against spatial disorder was confirmed by a field-theoretic renormalization group
calculation \cite{Janssen97}. This study did not produce a new critical fixed point but
runaway flow towards large disorder, suggesting unconventional behavior \footnote{Some
special versions of spatial disorder appear be lead to conventional behavior. A recent
computational study of the contact process on a Voronoi triangulation (a random lattice
with coordination disorder only) found critical behavior compatible with the \emph{clean}
DP universality class \cite{OAFD08}.}. Early Monte-Carlo simulations in two space
dimensions \cite{MoreiraDickman96,DickmanMoreira98} found logarithmically slow dynamics
in violation of power-law scaling. Moreover, in analogy with Griffiths singularities
\cite{Griffiths69}, very slow dynamics was found in an entire parameter region near the
transition \cite{Noest88,BramsonDurrettSchonmann91,WACH98,CafieroGabrielliMunoz98}.

A crucial step towards resolving this puzzling situation was taken by Hooyberghs et al.\
\cite{HooyberghsIgloiVanderzande03,HooyberghsIgloiVanderzande04} who used the Hamiltonian
formalism \cite{ADHR94} to map the one-dimensional disordered contact process onto a
random quantum spin chain. They then applied a Ma-Dasgupta-Hu strong-disorder
renormalization group \cite{MaDasguptaHu79,IgloiMonthus05} and showed that the transition
is controlled by an exotic infinite-randomness fixed point in the universality class of
the random transverse-field Ising model \cite{Fisher92,Fisher95}, at least for
sufficiently strong disorder. This type of critical point displays ultraslow activated
rather than power-law dynamical scaling. For weaker disorder, Hooyberghs et al.\ relied
on numerical simulations and predicted non-universal continuously varying exponents, with
either power-law or exponential dynamical scaling. Using large-scale Monte-Carlo
simulations Vojta and Dickison \cite{VojtaDickison05} recently confirmed the
infinite-randomness scenario. Moreover, their critical exponent values agreed with the
predictions of the strong-disorder renormalization group (which can be solved exactly in
one dimension), and they were universal, i.e., independent of disorder strength.

In higher dimensions, a strong-disorder renormalization group can still be applied,
but, in contrast to one dimension, it cannot be solved
analytically. By implementing the renormalization group numerically, Motrunich et al.\
\cite{MMHF00} demonstrated the existence of an infinite-randomness fixed point in two
dimensions (and at least the flow towards larger disorder in three dimensions).
However, reliable estimates for the critical exponent values have been hard
to come by due to the small sizes available in the quantum simulations.

In this paper, we report the results of large-scale Monte-Carlo simulations of the
contact process on two-dimensional randomly diluted lattices. The purpose of this work is
twofold. First, we intend to determine beyond doubt the character and universality of the
critical behavior; is it of conventional power-law type, infinite-randomness type or an
even more exotic kind? Second, we intend to perform a careful data analysis that allows us
to compute reasonably accurate exponent values. These
will be important beyond the disordered contact process and apply to all systems
controlled by the same infinite-randomness fixed point including the two-dimensional
random transverse-field Ising magnet \cite{MMHF00}.

The paper is organized as follows. In section \ref{sec:model}, we introduce our model,
the contact process on a randomly diluted lattice. We then contrast the scaling theories for
conventional critical points and infinite-randomness critical points. We also summarize the
predictions for the Griffiths region. In section \ref{sec:mc}, we present our simulation
method and various numerical results together with a comparison to theory.
We conclude in section \ref{sec:conclusions} by summarizing our findings,
discussing three space dimensions, and relating our results to a broader
classification of phase transitions with quenched disorder \cite{Vojta06}.

\section{Model and phase transition scenarios}
\label{sec:model}

\subsection{Contact process on a diluted lattice}

The contact process \cite{HarrisTE74}, a prototypical system in the DP universality
class, can be interpreted as a simple model for the spreading of a disease. The clean
contact process is defined on a $d$-dimensional hypercubic lattice. Each lattice site
$\mathbf{r}$ can be active (infected) or inactive (healthy). In the course of the time
evolution, active sites can infect their neighbors or they can spontaneously heal (become
inactive). Specifically, the dynamics of the contact process is given by a
continuous-time Markov process during which active sites become inactive at a rate
$\mu$ while inactive site become active at a rate $\lambda n /(2d)$. Here, $n$ is the
number of active nearest neighbor sites. The infection rate $\lambda$ and the healing rate
$\mu$ (which can be set to unity without loss of generality)
are external parameters. Their ratio controls the behavior of the system.

For $\lambda \ll \mu$, healing dominates, and the absorbing state without any active
sites is the only steady state (inactive phase). For sufficiently large infection rate
$\lambda$, there is a steady state with a nonzero density of active sites (active phase).
These two phases are separated by a nonequilibrium phase transition in the DP universality
class at a critical infection rate $\lambda_c^0$.

We now introduce quenched spatial disorder into the contact process by randomly diluting
the underlying lattice. Specifically, we randomly remove each lattice site with
probability $p$ \footnote{We define $p$ is the fraction of sites removed rather than
the fraction of sites present.}.
 As long as the impurity concentration $p$ remains below the percolation
threshold $p_c$, the lattice has an infinite connected cluster of sites which can support
the active phase. Conversely, at dilutions above $p_c$, the lattice is decomposed into
disconnected finite-size clusters. Because the infection eventually dies out on any
finite cluster, there is no active phase for $p>p_c$. Consequently, the contact process
on a randomly diluted lattice has two different nonequilibrium transitions (see also Fig.\
\ref{fig:pd}), the generic
transition for $p<p_c$ which is driven by the dynamic fluctuations of the contact
process and a percolation transition at $p=p_c$ which is driven by the lattice geometry
\cite{VojtaLee06}. The two transitions are separated by a multicritical point which was
studied in Ref.\ \cite{DahmenSittlerHinrichsen07}.

The basic observable in the contact process is the average density of active sites
at time $t$,
\begin{equation}
\rho(t) = \frac 1 {L^d} \sum_{\mathbf{r}} \langle  n_\mathbf{r}(t) \rangle
\label{eq:rho_definition}
\end{equation}
where $n_\mathbf{r}(t)=1$ if the site $\mathbf{r}$ is active at time $t$ and $n_\mathbf{r}(t)=0$
if it is inactive. $L$ is the linear system size, and $\langle \ldots \rangle$ denotes the
average over all
realizations of the Markov process. The longtime limit of this density (i.e., the steady
state density)
\begin{equation}
\rho_{\rm stat} = \lim_{t\to\infty} \rho(t)
\label{eq:OP_definition}
\end{equation}
is the order parameter of the nonequilibrium phase transition.

\subsection{Conventional power-law scaling}
\label{subsec:power_law}

In this subsection, we briefly summarize the scaling theory for absorbing state
transitions controlled by conventional fixed points with power-law dynamical scaling
(see, e.g., Ref.\ \cite{Hinrichsen00}), using the clean contact process as an example.

When the critical point $\lambda_c$ is approached from the active phase, the order parameter
(steady state density) vanishes following the power law
\begin{equation}
\rho_{\rm stat} \sim (\lambda-\lambda_c)^\beta \sim \Delta^\beta
\end{equation}
where $\Delta=(\lambda-\lambda_c)/\lambda_c$ is the dimensionless distance from the
critical point, and $\beta$ is the order parameter critical exponent. In addition
to the average density, we also need to characterize the length and time scales of the
density fluctuations. When approaching the transition, the (spatial)
correlation length $\xi_\perp$ diverges as
\begin{equation}
\xi_\perp \sim |\Delta|^{-\nu_{\perp}}~.
\end{equation}
The correlation time $\xi_\parallel$ behaves like a power of the correlation length,
\begin{equation}
\xi_\parallel \sim \xi_\perp^z, \label{eq:powerlawscaling}
\end{equation}
i.e., the dynamical scaling is of power-law form. The three critical exponents
$\beta$, $\nu_\perp$ and $z$ completely characterize the directed percolation
universality class. The above relations allow us
to write down the finite-size scaling form of the density as a function of
$\Delta$, $t$, and $L$,
\begin{equation}
\rho(\Delta,t,L) = b^{\beta/\nu_\perp} \rho(\Delta b^{-1/\nu_\perp},t b^z, L b)~,
\label{eq:rho}
\end{equation}
where $b$ is an arbitrary dimensionless scale factor.

Two interesting observables can be studied if the time evolution starts from a single
active site in an otherwise inactive lattice. The survival probability $P_s(t)$ is
the probability that an active cluster survives at time $t$ when starting from such
a single-site seed at time 0. For directed
percolation, the survival probability scales exactly like the density
\footnote{At more general absorbing state transitions, e.g., with several
    absorbing states, the survival probability scales with an exponent $\beta'$ which may
    be different from $\beta$ (see, e.g., \cite{Hinrichsen00}).},
\begin{equation}
P_s(\Delta,t,L) = b^{\beta/\nu_\perp} P_s(\Delta b^{-1/\nu_\perp},t b^z, L b)~.
\label{eq:Ps}
\end{equation}
The pair connectedness function $C(\mathbf{r},t)=\langle
n_{\mathbf{r}}(t) \, n_{0}(0) \rangle$ describes the probability that site
$\mathbf{r}$ is active at time $t$ when starting from an initial condition with a
single active site at $\mathbf{r}=0$ and time $0$. Because $C$
involves a product of two densities, its scale dimension is $2\beta/\nu_\perp$, and the
full finite-size scaling form reads \footnote{This relation relies on hyperscaling; it is only valid below the
    upper critical dimension $d_c^+$, which is four for directed percolation}
\begin{equation}
C(\Delta,\mathbf{r},t,L) = b^{2\beta/\nu_\perp} C(\Delta b^{-1/\nu_\perp}, \mathbf{r}b, t
b^z, L b)~.
\label{eq:C}
\end{equation}
The total number of particles $N$ when starting from a single seed site can be obtained
by integrating the pair connectedness $C$ over all space. This leads to the scaling form
\begin{equation}
N(\Delta,t,L) = b^{2\beta/\nu_\perp - d} N(\Delta b^{-1/\nu_\perp},t b^z, L b)~.
\label{eq:N}
\end{equation}
Finally, the mean-square radius $R$ of the active cluster, being a length, scales as
\begin{equation}
R(\Delta,t,L) = b^{-1} R(\Delta b^{-1/\nu_\perp},t b^z, L b)~.
\label{eq:R}
\end{equation}

At the critical point, $\Delta=0$, and in the thermodynamic limit, $L\to\infty$, the
above scaling relations lead to the following predictions for the time dependencies of
observables: The density and the survival probability asymptotically decay like
\begin{equation}
\rho(t) \sim t^{-\delta}, \qquad P_s(t) \sim t^{-\delta}
\end{equation}
with $\delta=\beta/(\nu_\perp z)$. In contrast, the radius and the number of particles in a cluster
starting from a single seed site increase like
\begin{equation}
R(t) \sim t^{1/z}, \qquad N(t) \sim t^\Theta
\end{equation}
where $\Theta=d/z - 2\beta/(\nu_\perp z)$ is the so-called critical initial slip
exponent.

\subsection{Activated scaling at an infinite-randomness fixed point}
\label{subsec:activated}

In this subsection we summarize the scaling theory for an infinite-randomness fixed point
with activated scaling, as has been found in the one-dimensional disordered contact process
\cite{HooyberghsIgloiVanderzande03,VojtaDickison05,Vojta06}.

At an infinite-randomness fixed point, the dynamics is ultraslow. The power-law
dynamical scaling (\ref{eq:powerlawscaling}) gets replaced by activated scaling
\begin{equation}
\ln(\xi_\parallel/t_0) \sim \xi_\perp^\psi, \label{eq:activatedscaling}
\end{equation}
characterized by a new exponent $\psi$ called the tunneling exponent. (This name stems
from the random transverse-field Ising model where this type of scaling was first found.)
Here $t_0$ is a nonuniversal microscopic time scale.
The exponential relation between time and length
scales implies that the dynamical exponent $z$ is formally infinite. In contrast, the
static scaling behavior remains of power law type.

Another important characteristic of an infinite-randomness fixed point is that the
probability distributions of observables become extremely broad. Therefore, the average and typical
value of an observable do not necessarily agree because averages may be dominated by rare events.
Nonetheless, the scaling form of the \emph{average} density at the infinite-randomness critical
point is obtained by simply
replacing the power-law scaling combination $t b^z$ by the activated combination $\ln(t/t_0)
b^\psi$ in the argument of the scaling function:
\begin{equation}
\rho(\Delta,\ln(t/t_0),L) = b^{\beta/\nu_\perp} \rho(\Delta b^{-1/\nu_\perp},\ln(t/t_0) b^\psi, L
b)~. \label{eq:rho_activated}
\end{equation}
Analogously, the scaling forms of the average survival probability, the average number
of sites in a cluster, and the mean-square cluster radius  starting from a single site are
\begin{equation}
P_s(\Delta,\ln(t/t_0),L) = b^{\beta/\nu_\perp} P_s(\Delta b^{-1/\nu_\perp},\ln(t/t_0) b^\psi,L
b) \label{eq:Ps_activated}
\end{equation}
\begin{equation}
 N(\Delta,\ln(t/t_0),L) = b^{2\beta/\nu_\perp -d} N(\Delta
b^{-1/\nu_\perp},\ln(t/t_0) b^\psi,L b).  \label{eq:N_activated}
\end{equation}
\begin{equation}
 R(\Delta,\ln(t/t_0),L) = b^{-1} R(\Delta
b^{-1/\nu_\perp},\ln(t/t_0) b^\psi,L b).  \label{eq:R_activated}
\end{equation}

These activated scaling forms lead to logarithmic time dependencies at the critical point
(in the thermodynamic limit). The average density and the survival probability
asymptotically decay like
\begin{equation}
\rho(t) \sim [\ln(t/t_0)]^{-\bar\delta}, \qquad P_s(t) \sim [\ln(t/t_0)]^{-\bar\delta}
\label{eq:logdecay}
\end{equation}
with $\bar\delta=\beta/(\nu_\perp \psi)$ while the radius and the average number of particles in a
cluster starting from a single seed site increase like
\begin{equation}
R(t) \sim [\ln(t/t_0)]^{1/\psi}, \qquad N(t) \sim [\ln(t/t_0)]^{\bar\Theta} \label{eq:clustersize}
\end{equation}
with $\bar\Theta=d/\psi-2\beta/(\nu_\perp \psi)$.

In contrast to one dimension, where the critical exponents can be calculated exactly
within the strong-disorder renormalization group,  they need to be
found numerically in higher dimensions.

\subsection{Griffiths singularities}
\label{subsec:Griffiths}

Quenched spatial disorder does not only destabilize the DP critical point, it also leads
to interesting singularities, the so-called Griffiths singularities \cite{Griffiths69},
in an entire parameter region around the transition. For a recent review of this topic
at thermal, quantum, and nonequilibrium transitions, see Ref.\ \cite{Vojta06}.
In the context of the contact process on a diluted lattice, Griffiths singularities can
be understood as follows.

Because lattice dilution reduces the tendency towards the active phase, the critical
infection rate of the diluted system is higher than that of the clean system, $\lambda_c
> \lambda_c^0$. For infection rates $\lambda$ with $\lambda_c > \lambda > \lambda_c^0$, the system is
globally in the inactive phase,
i.e., it eventually decays into the absorbing state. However, in the
thermodynamic limit, one can find arbitrarily large spatial regions devoid of impurities.
These rare regions are locally in the active phase. They cannot support a
non-zero steady state density because they are of finite size, but their decay is very slow
because it requires a rare, exceptionally large density fluctuation.

As a result, the inactive phase of the contact process on a diluted lattice can be divided
into two regions. For infection rates below the clean critical point, $\lambda<\lambda_c^0$,
the behavior is conventional. The system
approaches the absorbing state exponentially fast in time. The decay time increases with
$\lambda$ and diverges as $|\lambda-\lambda_c^0|^{-z\nu_\perp}$ where $z$ and $\nu_\perp$
are the exponents of the clean critical point \cite{VojtaDickison05,DickisonVojta05}.
Inside the Griffiths region, $\lambda_c > \lambda \ge \lambda_c^0$, one has to estimate
the rare region contribution to the time evolution of the density \cite{Noest86,Noest88}.
The probability $w$ for finding a rare
region of linear size $L_r$ devoid of impurities is (up to pre-exponential factors) given
by
\begin{equation}
 w(L_r) \sim \exp( -\tilde p L_r^d)
\end{equation}
with $\tilde p = - \ln(1-p)$. To exponential accuracy, the rare region contribution to the density
can be written as
\begin{equation}
\rho(t) \sim \int dL_r ~L_r^d ~w(L_r) \exp\left[-t/\tau(L_r)\right] \label{eq:rrevo}
\end{equation}
where $\tau(L_r)$ is the decay time of a rare region of size $L_r$. Let us first discuss
the behavior at the clean critical point, $\lambda_c^0$, i.e., at the boundary between
the conventional inactive phase and the Griffiths region. Here, the decay time
of a single, impurity-free rare region of size $L_r$ scales as $\tau(L_r) \sim L_r^z$
with $z$ the clean critical exponent [see (\ref{eq:rho})].
Using the saddle point method to evaluate the integral (\ref{eq:rrevo}), we find the
leading long-time decay of the density to be given by a stretched exponential,
\begin{equation}
\ln \rho(t) \sim - \tilde{p}^{z/(d+z)}~ t^{d/(d+z)}~, \label{eq:stretched}
\end{equation}
rather than a simple exponential decay as for $\lambda<\lambda_c^0$.

Within the Griffiths region, $\lambda_c^0<\lambda<\lambda_c$, the decay time of
a single rare region depends exponentially on its volume,
\begin{equation}
\tau(L_r) \sim \exp(a L_r^d)
\end{equation}
because a coordinated fluctuation of the entire rare region is required to take it to the
absorbing state \cite{Noest86,Noest88,Schonmann85}. The nonuniversal prefactor $a$
vanishes at the clean critical point $\lambda_c^0$ and increases with $\lambda$. Close to
$\lambda_c^0$, it behaves as $a \sim \xi_\perp^{-d} \sim
(\lambda-\lambda_c^0)^{d\nu_\perp}$ with $\nu_\perp$ the clean critical exponent.
Repeating the saddle point analysis of the integral (\ref{eq:rrevo}) for this case, we
obtain a power-law decay of the density
\begin{equation}
\rho(t) \sim  t^{-\tilde p/a}  = t^{-d/z'} \label{eq:griffithspower}
\end{equation}
where $z'=da/\tilde p$ is a customarily used nonuniversal dynamical exponent in the
Griffiths region. Its behavior close to the \emph{dirty} critical point $\lambda_c$ can
be obtained within the strong disorder renormalization group method
\cite{HooyberghsIgloiVanderzande04,Fisher95,MMHF00}. When approaching the phase transition, $z'$ diverges as
\begin{equation}
z' \sim  |\lambda-\lambda_c|^{-\psi\nu_\perp}
\label{eq:z_prime}
\end{equation}
where $\psi$ and $\nu_\perp$ are the
exponents of the dirty critical point.

Let us emphasize that strong power-law Griffiths singularities such as (\ref{eq:griffithspower})
usually occur in connection with infinite-randomness critical points, while conventional
critical points display exponentially weak Griffiths effects \cite{Vojta06}. Thus, the
character of the Griffiths singularities can be used to identify the correct scaling scenario.

\section{Monte-Carlo simulations}
\label{sec:mc}

\subsection{Method and overview}

We now turn to the main objective of our work, large-scale Monte-Carlo simulations of the
contact process on a randomly diluted square lattice. There are several efficient
computational implementations of the contact process, all equivalent with respect to the
universal behavior at the phase transition. We followed the algorithm described, e.g., by
Dickman \cite{Dickman99}. The simulation starts at time $t=0$ from some configuration of
active and inactive sites. Each event consists of randomly selecting an active site
$\mathbf{r}$ from a list of all $N_a$ active sites, selecting a process: creation with
probability $\lambda/(1+ \lambda)$ or annihilation with probability $1/(1+ \lambda)$ and,
for creation, selecting one of the four neighboring sites of $\mathbf{r}$. The creation
succeeds, if this neighbor is empty (and not an impurity site). The time increment
associated with this event is $1/N_a$.

Using this method, we investigated systems with sizes of up to $8000 \times 8000$
sites and impurity (vacancy) concentrations $p=0$, 0.1, 0.2, 0.3, and $p=p_c=0.407254$.
To explore the ultraslow dynamics predicted in the infinite-randomness scaling scenario,
we simulated very long times up to $t=10^{10}$ which is, to the best of our
knowledge, significantly longer than all previous simulations of the disordered
two-dimensional contact process. In all cases, we averaged over many disorder
realizations, details will be mentioned below for each specific simulation.  The total
numerical effort was approximately 40000 CPU-days on the Pegasus Cluster at Missouri
S\&T.

\begin{figure}[tb]
\centerline{\includegraphics[width=8cm]{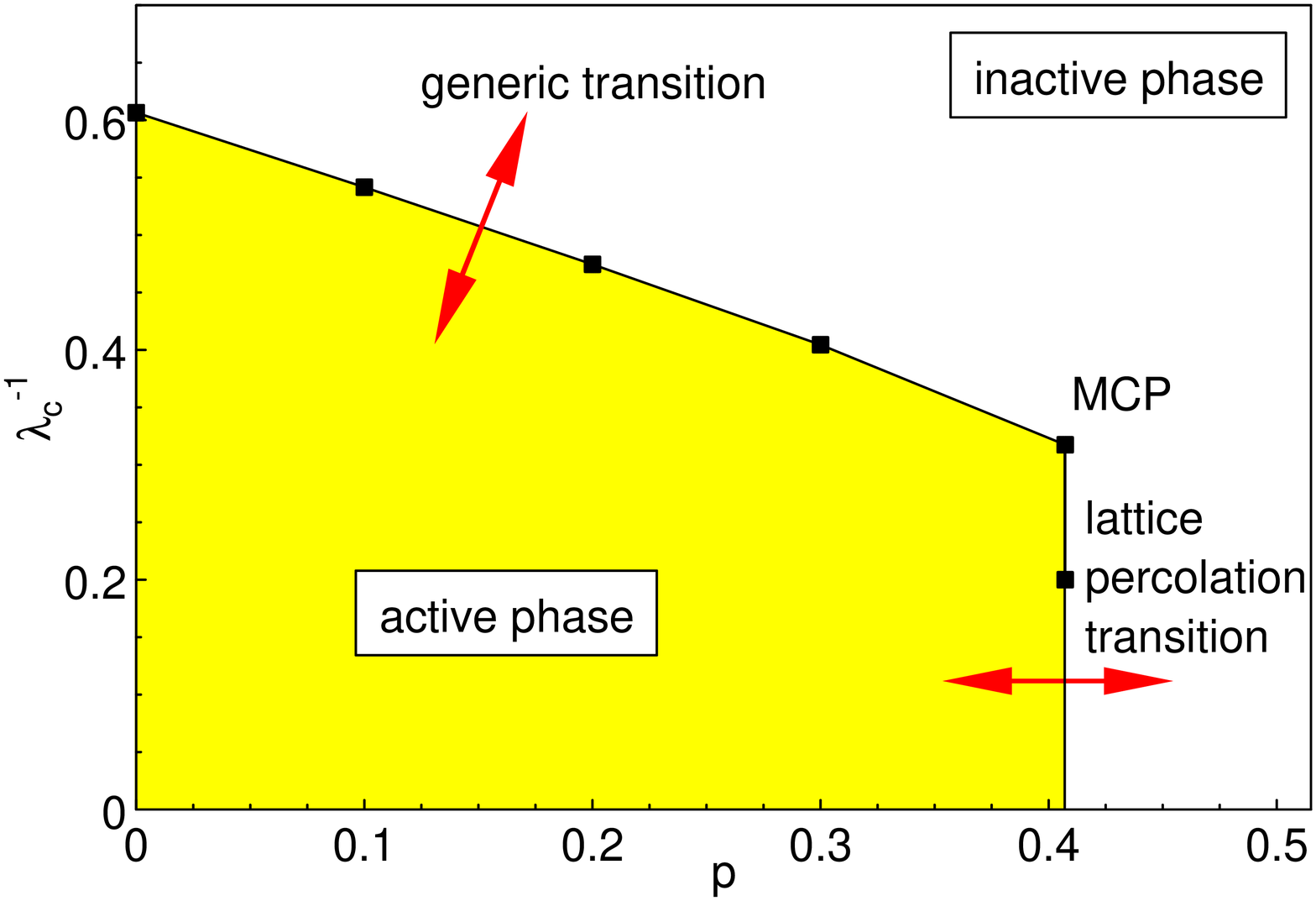}}
\caption{(Color online) Phase diagram of the contact
process on a site-diluted square lattice
         (inverse critical infection rate $\lambda_c^{-1}$ vs impurity concentration $p$).
         MCP marks the multicritical point. The
          black dots show the actual simulation results, the lines are guides to
          the eye.}
\label{fig:pd}
\end{figure}

To identify the critical point for each parameter set, and to determine the critical
behavior, we carried out two types of simulations, (i) runs starting from a
completely active lattice during which we monitor the time evolution of the density
$\rho(t)$, and (ii) runs starting from a single active site in an otherwise inactive
lattice during which we monitor the survival probability $P_s(t)$ and the size $N_s(t)$
of the active cluster. Figure \ref{fig:pd} gives an overview of the phase diagram
resulting from these simulations. As expected, the critical infection rate $\lambda_c$
increases with increasing impurity concentration. In the following subsections we discuss
the behavior in the vicinity of the phase transition in more detail.

\subsection{Clean 2d contact process}
\label{subsec:clean}

\begin{figure}[tb]
\centerline{\includegraphics[width=8cm]{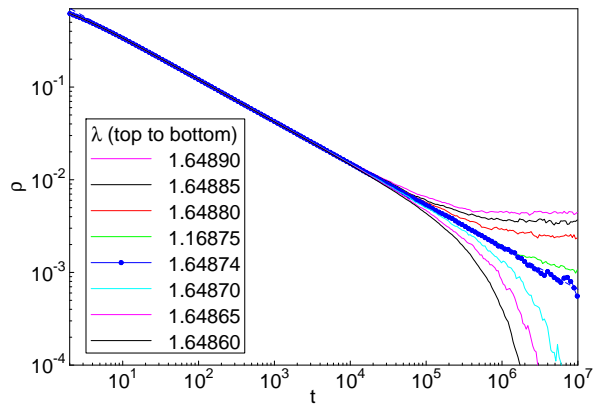}}
\caption{(Color online) Time evolution of the average density $\rho(t)$ for the contact process on an
         undiluted ($p=0$) square lattice, starting from a completely active lattice.}
\label{fig:cleanrho}
\end{figure}

We first performed a number of simulations for clean undiluted lattices ($p=0$),
mainly to test our numerical implementation of the contact process. Fig.\
\ref{fig:cleanrho} shows the density $\rho(t)$ for runs starting from a completely
active lattice. The data are averages over 60 runs of a system of linear size $L=4000$
except for the two infection rates closest to the transition, $\lambda=1.64874$ and
1.64875 for which we performed 42 runs of a system of size $L=8000$.
From these simulations, we estimate the critical infection rate to be
$\lambda_c^0=1.64874(2)$ where the number in brackets gives the error of the last digit.
The critical exponent $\delta$ can be determined from a power-law fit of $\rho(t)$;
we find $\delta=0.4526(7)$.

We then perform a scaling analysis of $\rho(t)$ [based on (\ref{eq:rho})]
on the inactive side of the transition using 12 different infection rates between
1.60 and $\lambda_c^0=1.64874$. This allows us to find the exponent combination $z
\nu_\perp = 1.290(4)$. Finally, we carried out several runs right at the critical
point but with system sizes varying from $L=25$ to $L=8000$. A finite-size scaling
analysis [again based on (\ref{eq:rho})] gives the dynamical exponent
$z=1.757(8)$. All other critical exponents can now be calculated from the scaling
relations (\ref{eq:rho}), (\ref{eq:Ps}), and (\ref{eq:N}). We find $\nu_\perp=0.734(6)$,
$\beta=0.584(3)$, and $\Theta=0.233(6)$.

To verify the results we also carried out simulations starting from a single
active seed site. Using 500,000 runs with a maximum time of $10^6$, we arrived at
the same value $\lambda_c^0=1.64874(2)$ for the critical infection rate. We measured
the survival probability $P_s(t)$, the number of active sites $N_s(t)$ and the mean
square radius of the active cluster $R(t)$. The resulting estimates of
the exponents $\delta$, $\Theta$, and $z$ confirmed the values quoted above.

Our results are in excellent
agreement with, and of comparable accuracy to, other large-scale simulation of the DP
universality class in two dimensions (see, e.g., Ref.\ \cite{Dickman99} and references
therein).

\subsection{Diluted 2d contact process: Overview}
\label{subsec:overview}

\begin{figure}[tb]
\centerline{\includegraphics[width=8cm]{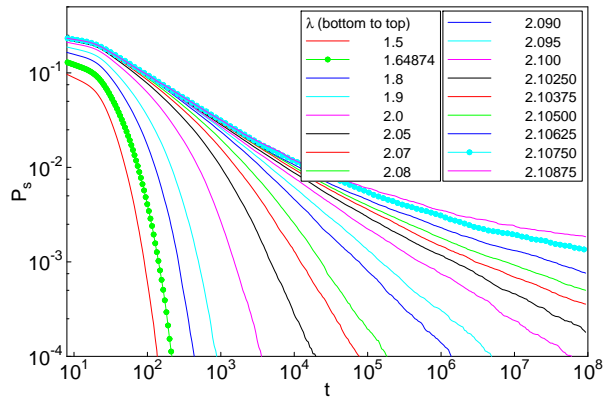}}
\caption{(Color online) Time evolution of the survival probability $P_s(t)$ for impurity concentration
         $p=0.2$, starting from a single seed site. The clean critical infection rate $\lambda_c^0=1.64874$
         and the actual critical point $\lambda_c=2.10750$ are specially marked.}
\label{fig:Psoverview}
\end{figure}

We start our discussion of the diluted contact process by showing in
Fig. \ref{fig:Psoverview} an overview of the survival probability $P_s(t)$ for a
system with impurity concentration $p=0.2$, covering the $\lambda$ range from the
conventional inactive phase, $\lambda<\lambda_c^0$, all the way to the active phase,
$\lambda > \lambda_c$.
The data represent averages over 5000 disorder configurations, with 128 runs for each disorder configuration
starting from random seed sites. The system size, $L=2000$, is
chosen such that the active cluster stays much smaller then the sample over the time
interval of the simulation (thus eliminating finite-size effects).

For infection rates below and at the clean critical point $\lambda_c^0=1.64874$, the
density decay is very fast, clearly faster than a power law. Above $\lambda_c^0$, the
decay becomes slower and asymptotically appears to follow a power law with an exponent
that varies continuously with $\lambda$. For even larger infection rates, the decay
seems to be slower than a power law. At the largest infection rates, the system will
ultimately reach a nonzero steady-state survival probability, i.e., the it is in the active phase.

Let us emphasize that the behavior in Fig.\ \ref{fig:Psoverview}, in particular, the nonuniversal
power-law decay over a range of infection rates, already provides support for the infinite-randomness scenario
of subsection \ref{subsec:activated} while it disagrees with the conventional
scenario (subsection \ref{subsec:power_law}) of power-law behavior restricted to the critical point
and exponentially weak Griffiths effects.

\subsection{Finding the critical point}
\label{subsec:critical_point}

\begin{figure}[tb]
\centerline{\includegraphics[width=8.cm]{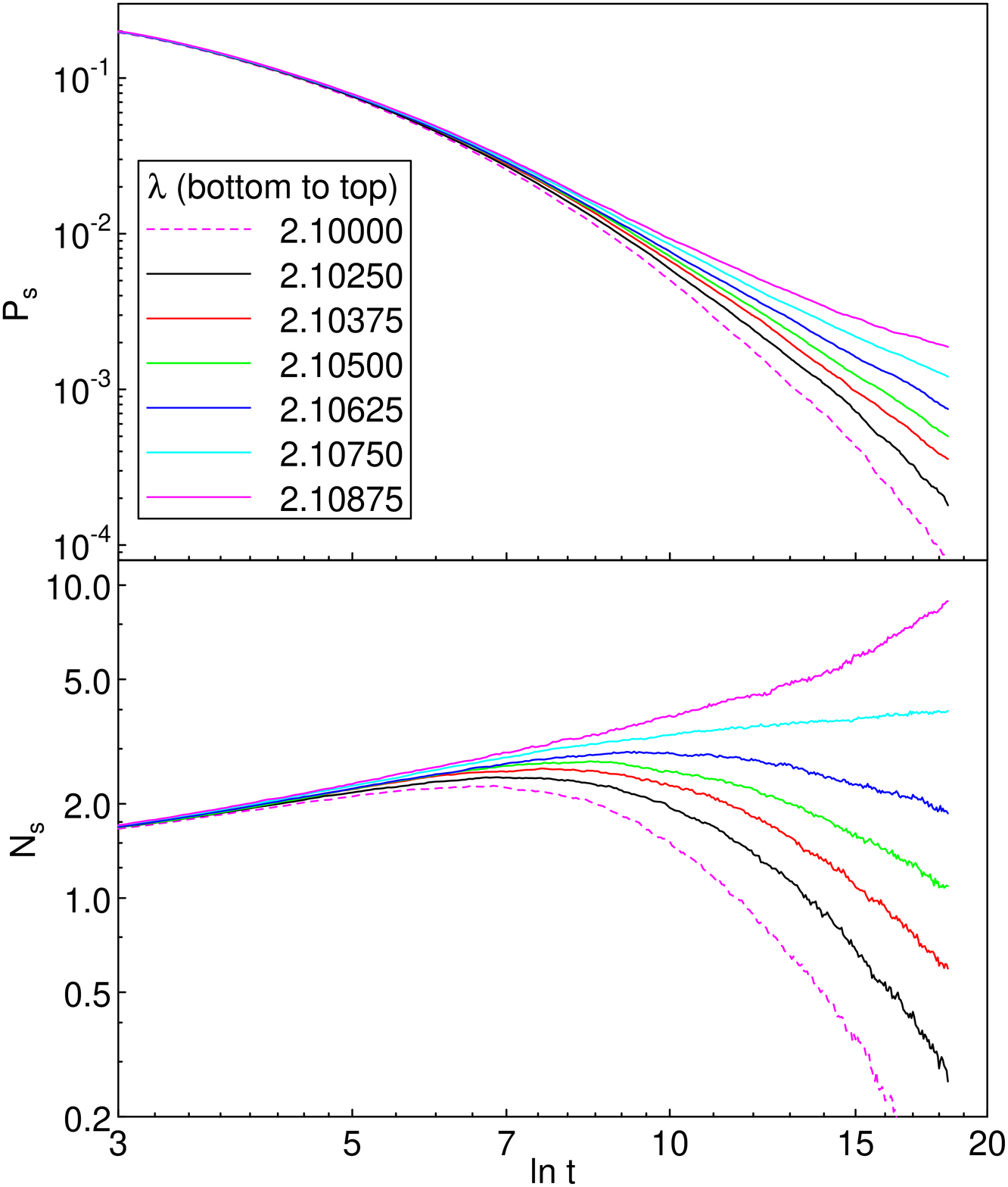}}
\caption{(Color online) $\ln P_s(t)$ and $\ln N_s(t)$ versus $\ln \ln t$ for several infection rates
         close to the critical point (at least 5000 disorder realizations with 128 trials each,
         $p=0.2$).}
\label{fig:criticalp02}
\end{figure}

A very efficient method for identifying the critical point in the clean contact process
and other clean reaction-diffusion systems is to look for the critical power-law time
dependencies of various observables. To use the same idea in our diluted contact process
simulations, we plot $\ln P_s(t)$, $\ln N_s(t)$ and $\ln \rho(t)$ versus $\ln \ln t$.
In such plots, the anticipated logarithmic laws (\ref{eq:logdecay}) and (\ref{eq:clustersize})
correspond to straight lines. Figure \ref{fig:criticalp02} shows these plots for
several infection rates close to the critical point.
The parameters are as in subsection \ref{subsec:overview}. At the first glance, the data
at the appropriate critical infection rate appear to follow the expected
logarithmic behavior over several orders of magnitude in time $t$. However, a closer
inspection reveals a serious problem. The survival probability $P_s$ suggests a critical
infection rate $\lambda_c$ in the range 2.10500 to 2.10625 while the cluster size
$N_s$ appears subcritical even at the larger value 2.10750. Such a discrepancy suggests
strong corrections to scaling.

The origin of these corrections can be easily understood by rewriting $P_s(t) \sim
[\ln(t/t_0)]^{-\bar\delta} = [\ln(t) -\ln(t_0)]^{-\bar\delta}$. The microscopic time
scale thus provides a correction to scaling; and since we cover only a moderate range
in $\ln(t)$, it strongly influences the results. We emphasize that this problem is intrinsic
to the case of activated scaling (and thus unavoidable) because logarithms, in
contrast to power laws, are not scale free. Identifying straight lines in plots like
Fig. \ref{fig:criticalp02} is thus not a reliable tool for finding the critical
point, even more so because simulations in 1D (where the analysis is much easier because
the critical exponents are known exactly) have given sizable $t_0 \approx 100 \ldots 500$
for typical parameter values \cite{VojtaDickison05}.

To circumvent this problem, we devised a method for finding the critical point
{\em without} needing a value for the microscopic scale $t_0$. It is based on the
observation that in the infinite-randomness scenario, $t_0$ has the same value in the
scaling forms of all observables (because it is related to the basic energy scale of
the strong-disorder renormalization group). Thus, if we plot $N_s(t)$ versus $P_s(t)$,
the critical point corresponds to power-law behavior provided all other corrections
to scaling are weak. The same is true for other combinations of observables.

\begin{figure}[tb]
\centerline{\includegraphics[width=8.cm]{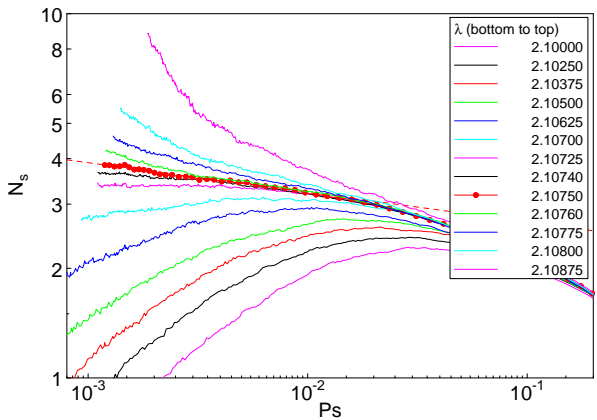}}
\caption{(Color online) $\ln N_s(t)$ versus $\ln P_s(t)$ and  for several infection rates
         close to the critical point. The straight line is a power-law fit of
         the asymptotic part of the $\lambda=2.10750$ curve.
         ($p=0.2$, 5000 - 20,000 disorder realizations with
         128 trials each.}
\label{fig:ns_vs_ps}
\end{figure}

Figure \ref{fig:ns_vs_ps} shows a plot of $N_s(t)$ versus $P_s(t)$ for several infection rates
close to the phase transition.
The figure  shows that the data for $\lambda=2.1075$ follow the expected power law
for about a decade in $P_s$ ($10^{-3}<P_s<10^{-2}$, corresponding to 4 decades of
time, from about
$10^4$ to $10^8$), while the data at higher and lower $\lambda$ curve away as
expected. We therefore identify $\lambda_c=2.1075(1)$ as the critical infection rate for
dilution $p=0.2$. This value agrees well with an estimate
based on Dickman's \cite{MoreiraDickman96} heuristic criteria of $\lambda_c$
being the smallest $\lambda$ supporting asymptotic growth of $N_s(t)$.

Figure \ref{fig:ns_vs_ps} also reveals two remarkable features of the
problem. (i) The asymptotic critical region is extremely narrow,
$|\Delta|=|\lambda-\lambda_c|/\lambda_c \lessapprox 10^{-3}$,
at least for these two observables.
This follows from fact that the data for $\lambda$ outside the small
interval [2.1070, 2.1080] curve away from the critical
curve \emph{well before} it reaches the asymptotic regime. (ii) The crossover to
the asymptotic behavior occurs very late since $P_s \approx 0.01$ corresponds to
a crossover time of $t_x \approx 10^4$, implying that very long simulations are
necessary to extract the true critical behavior. Moreover, the \emph{mean-square}
radius of a surviving active cluster at the crossover time is $R(t_x) \approx 55$,
corresponding to an overall diameter of about 200. This means simulations with linear
system sizes below about 200 will never reach the asymptotic behavior. We confirmed
this observation by carrying out simulations with smaller sizes.

Having determined the critical point, we can now obtain a rough estimate of the microscopic
time scale $t_0$ by replotting $\ln P_s(t)$ and $\ln N_s(t)$ as functions of $\ln \ln(t/t_0)$
with varying $t_0$. At the correct value for $t_0$, the critical curves ($\lambda=2.1075$)
of {\em all} observables must be straight lines.  By performing this analysis for $P_s(t)$,
$N_s(t)$, and $\rho(t)$ we find $\ln t_0 = 5.5 \pm 1.0$ for dilution
$p=0.2$. The results for all the
observables agree within the error bars, providing an additional consistency check for
our method and the underlying infinite-randomness scenario.

\subsection{Critical behavior}
\label{subsec:critical_behavior}

According to subsection \ref{subsec:activated}, the critical behavior at the
infinite-randomness fixed point can be characterized by three independent exponents, e.g.,
$\beta$, $\nu_\perp$, and $\psi$. The two ``static'' exponents $\beta$ and
$\nu_\perp$ can in principle be determined without reference to $t_0$, the
value of the tunneling exponent $\psi$ depends on the above estimate for $t_0$.

The combination $\beta/\nu_\perp$ (the scale dimension of the order parameter) can be
obtained directly from the critical curve in
Fig.\ \ref{fig:ns_vs_ps}. By combining eqs.\ (\ref{eq:logdecay}) and
(\ref{eq:clustersize}), we find $N_s \sim P_s^{-\bar\Theta/\bar\delta} \sim
P_s^{2(1-\nu_\perp/\beta)}$. A fit of the asymptotic part of the critical curve gives
$\beta/\nu_\perp =0.96(2)$. Here, the error is almost entirely due to the uncertainty
in locating the critical infection rate $\lambda_c$, the statistical error is much
smaller. We obtained independent estimates for $\beta/\nu_\perp$ from analyzing plots
of $N_s$ vs. $R$ and $P_s$ vs. $R$ in a similar fashion. The results agree with the above
value within their error bars.

\begin{figure}[tb]
\centerline{\includegraphics[width=8.0cm]{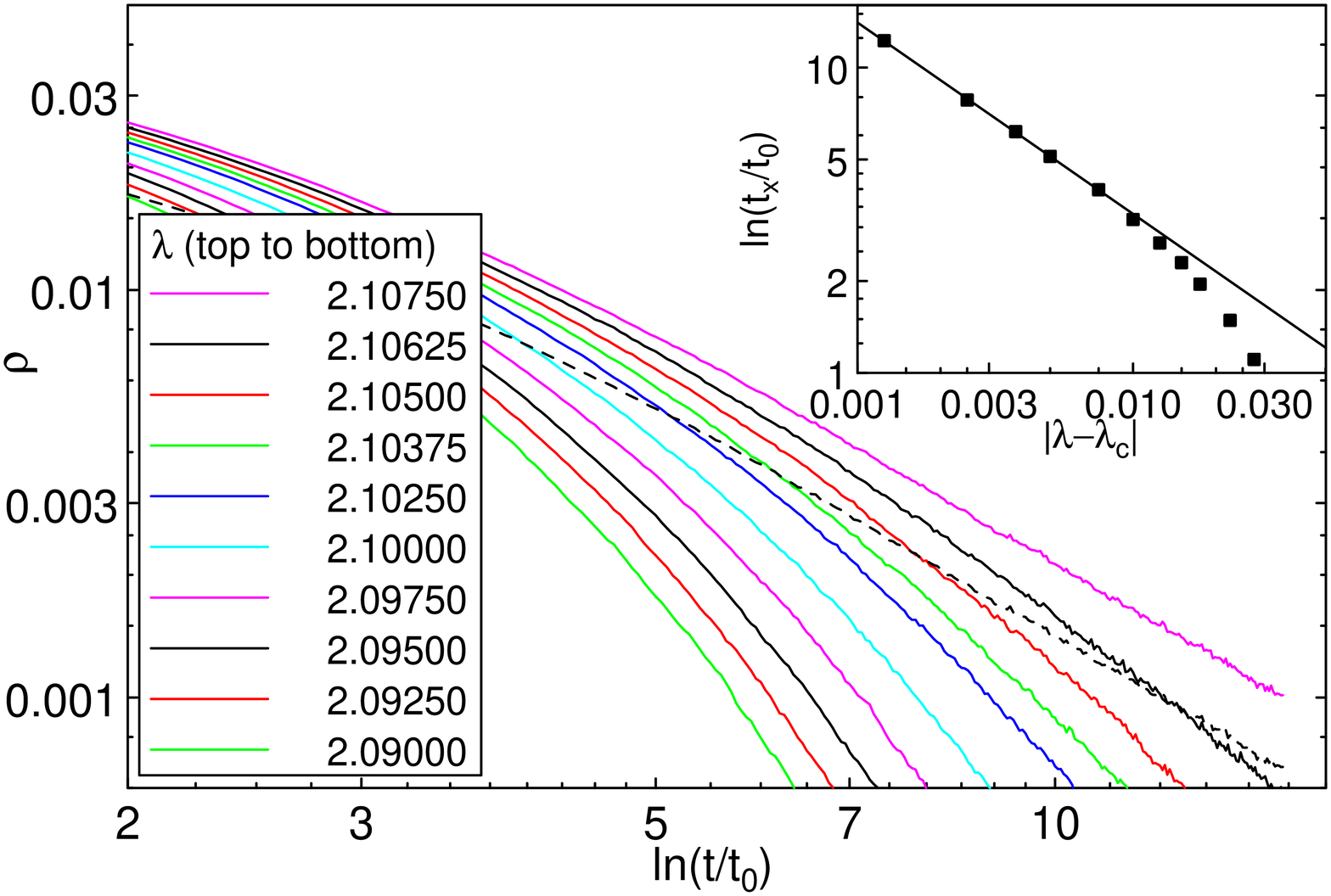}}
\caption{(Color online) $\ln \rho(t)$ versus $\ln \ln(t/t_0)$ and  for several infection rates
         at and below to the critical point. The dashed line represents
         2/3 of the critical $\rho$. Inset: Time $t_x$ at which the
         density falls below 2/3 of the critical value as a function
         of the distance from criticality. The solid line is a power-law
         fit of $\ln(t_x/t_0)$ vs $\Delta$.
         ($p=0.2$, 13 disorder realizations with $L=8000$ for $\lambda>2.1$,
         250 realizations with $L=2000$ for $\lambda \le 2.1$) }
\label{fig:nu-scaling}
\end{figure}

To obtain the tunneling exponent $\psi$, we now fit the critical survival probability
$P_s(t)$ to $[\ln(t/t_0)]^{-\bar\delta}$ (for times up to $t=10^8$) using our estimate
$\ln t_0 =5.5\pm 1.0 $. We find $\bar\delta=1.9(2)$ with the error mainly coming from the
uncertainties in $\lambda_c$ and $t_0$. An analogous analysis of the density of active
sites $\rho(t)$ in a simulation starting from a fully infected lattice ($L=8000$, times up to
$t=10^9$) gives exactly the same value (see Fig.\ \ref{fig:nu-scaling}).
From the time evolution of $N_s(t)$, we also find the estimate $\bar \Theta=0.15(3)$
for the initial slip exponent.  Using $\psi = \beta/(\nu_\perp \bar\delta)$
gives the tunneling exponent, $\psi=0.51(6)$. Estimates obtained from $N_s(t)$ and directly from fitting $R(t)$ to
$[\ln(t/t_0)]^{1/\psi}$ agree within the error bars with this value. It must be
emphasized that the value of $\psi$ sensitively depends on the correct identification of
$t_0$. A naive fit of $P_s(t)$ to a power of $\ln(t)$ would give  a value of about 3.0 for $\bar \delta$ and
thus a much smaller value of about 0.33 for $\psi$.

Finding the final missing exponent in our triple ($\beta$, $\nu_\perp$, $\psi$) requires
the analysis of off-critical simulation runs. Here, the extreme narrowness of the
asymptotic critical region causes serious problems, because we have only three $\lambda$
values from 2.1070 to $\lambda_c=2.1075$. Moreover, a glance at Fig.\  \ref{fig:ns_vs_ps}
shows that adding simulations runs at additional $\lambda$ closer to the critical point
would only
make sense if we also significantly increased the simulation time and/or the number of
disorder realizations. Because this appears to be beyond our present computational
capabilities, we determine an estimate for $\beta$ by extrapolating the effective
exponent obtained outside the asymptotic critical region.
To this end, we analyze the time evolution of the density $\rho(t)$ starting from a
fully infected lattice show in Fig.\  \ref{fig:nu-scaling} for various infection
rates $\lambda$ below the critical point. For each $\lambda$, we define a crossover
time $t_x$ at which the density $\rho(t)$ is exactly 2/3 of the critical density at
the same time.

\begin{figure}[tb]
\centerline{\includegraphics[width=8.cm]{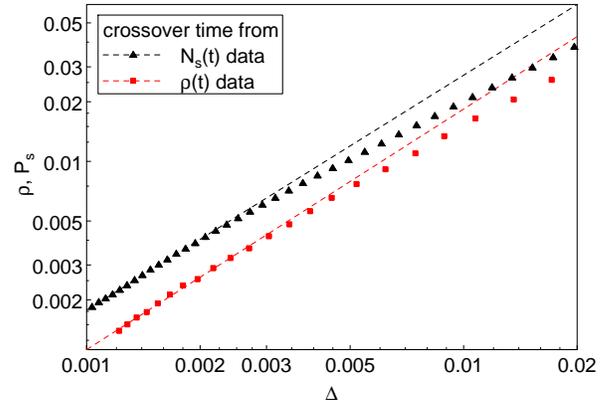}}
\caption{(Color online) $\ln\rho(t)$ vs. $\ln\Delta(t)$ at the crossover time $t=t_x$ for the density itself and
         $\ln P_s(t)$ vs. $\ln\Delta(t)$ at the crossover time $t=t_x$ for the particle number $N_s$.
         The dashed lines are power-law fits to the expected behavior $\rho \sim P_s \sim |\Delta|^\beta$.}
\label{fig:cross_cmp}
\end{figure}

This crossover time can be analyzed in two ways. According to (\ref{eq:rho_activated}),
it should behave as $\ln(t_x/t_0) \sim |\Delta|^{-\nu_\perp\psi}$. The inset of Fig.\  \ref{fig:nu-scaling}
shows a fit to this form, giving $\nu_\perp\psi = 0.64(10)$. However, this value depends on
our estimate for the microscopic scale $t_0$. To avoid relying on $t_0$, we plot
$|\Delta(t_x)|$ vs. the critical $\rho(t)$ (at $t=t_x$) in Fig.\ \ref{fig:cross_cmp}.
According to (\ref{eq:rho_activated}), the expected behavior is a power law $\rho \sim
|\Delta|^\beta$. The same analysis can also be performed by analyzing the crossover
of $N_s(t)$ and plotting the resulting $\Delta(t_x)$ vs. $P_s(t)$. Both data sets show
significant curvature (i.e. deviations from the expected power law behavior) which is not
surprising because the underlying $\lambda$ are not in the asymptotic region.
By fitting the small-$\Delta$ part of both curves, we estimate $\beta= 1.15(15)$.
A fit of  $\Delta(t_x)$ vs. $N_s(t)$ gives $\nu_\perp-\beta=0.045(15)$. Thus, our final
estimate for the correlation length exponent $\nu_\perp=1.20(15)$ in agreement with the
Harris criterion $\nu_\perp \ge 2/d = 1$.

\subsection{Universality}

The universality of the infinite-randomness scenario for the critical point in the
disordered contact process has been controversially discussed in the past
\cite{HooyberghsIgloiVanderzande04,VojtaDickison05,NeugebauerFallertTaraskin06}. The
underlying strong-disorder renormalization group becomes exact only in the limit of broad
disorder distributions and can thus not directly address the fate of a weakly disordered
system. However, Hoyos \cite{Hoyos08} recently showed that within this renormalization
group scheme, the disorder always increases under renormalization, even if it is weak
initially. Moreover, the perturbative renormalization group of Janssen \cite{Janssen97},
which is controlled for weak disorder, displays runaway flow towards large disorder
strength supporting a universal scenario in which the infinite-randomness fixed point
controls the transition for all bare disorder strength.

\begin{figure}[tb]
\centerline{\includegraphics[width=8.cm]{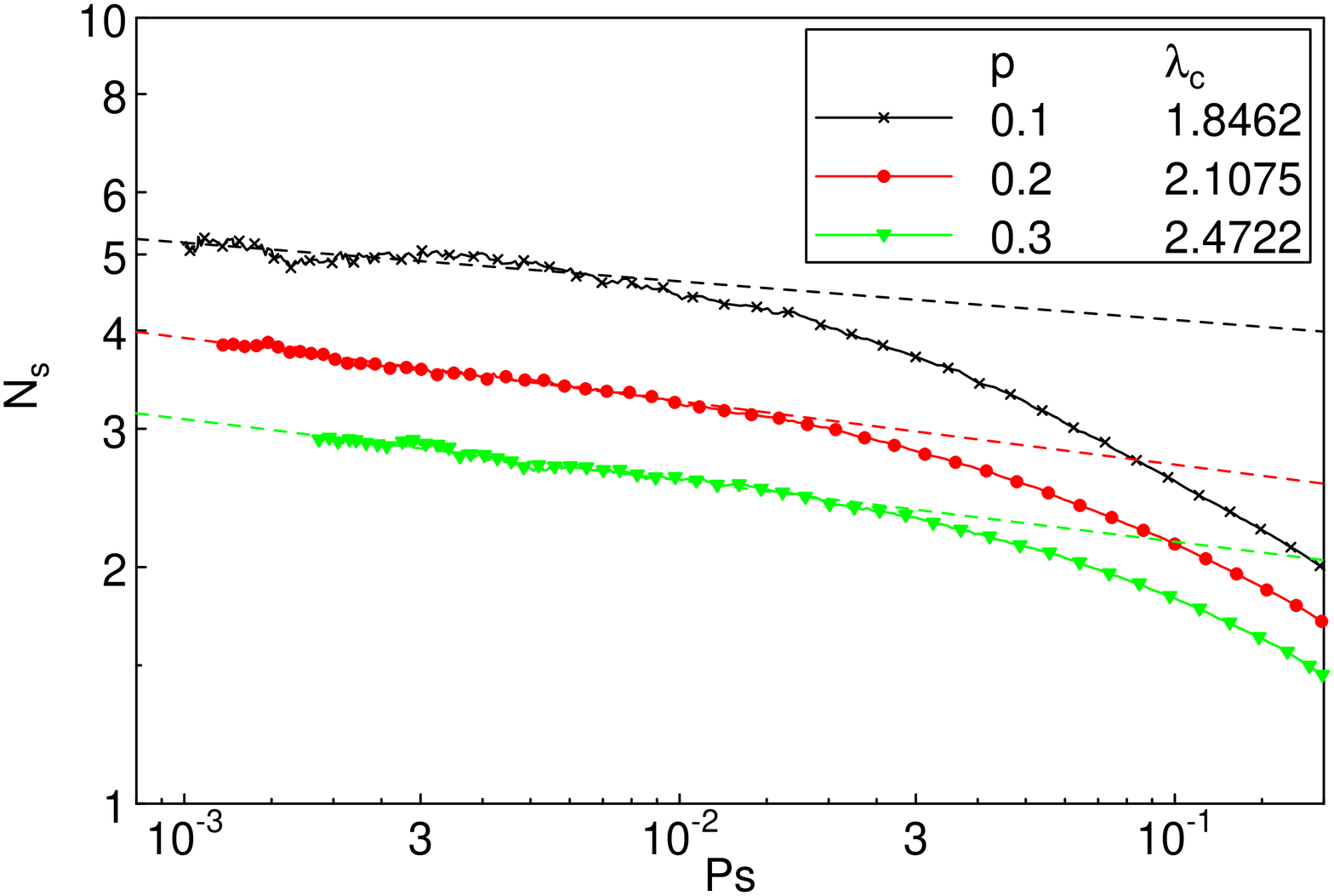}}
\caption{(Color online) $\ln N_s(t)$ versus $\ln P_s(t)$ at criticality for impurity concentrations
         $p=0.1$, 0.2, and 0.3. The dashed lines are power-law fits of the asymptotic
         parts (18000 disorder configurations for $p=0.2$, 4000 each for $p=0.1$ and 0.3,
         128 spreading trials for each configuration).}
\label{fig:all_critical}
\end{figure}

To address this question in our simulations, we repeat the above analysis for impurity
concentrations $p=0.1$ and 0.3 albeit with somewhat smaller numbers of disorder
realizations. We determine the critical infection rates from plots of $N_s(t)$ vs.
$P_s(t)$ analogous to Fig.\ \ref{fig:ns_vs_ps} and find $\lambda_c=1.8462(3)$ for
$p=0.1$ and $\lambda_c=2.4722(2)$ for $p=0.3$. Figure \ref{fig:all_critical} shows the
critical $N_s$ vs. $P_s$ curves for $p=0.1$, 0.2, and 0.3.
In the long-time (low-$P_s$) limit, all three curves follow power-laws, and the
resulting values for the exponent $\beta/\nu_\perp$ agree within their error bars,
$\beta/\nu_\perp=0.97(3)$ for $p=0.1$, $\beta/\nu_\perp=0.96(2)$ for $p=0.2$,
and $\beta/\nu_\perp=0.96(3)$ for $p=0.3$.

Having fixed the critical point for all $p$, we determine the microscopic time scale
$t_0$ as in subsection \ref{subsec:critical_point}. Its value changes significantly with
$p$, it is $\ln t_0 \approx 6.5$ (for $p=0.1$), 5.5 (for $p=0.2$), and 4.0 (for $p=0.3$).
Power-law fits of $P_s(t)$ and $N_s(t)$ to $\ln(t/t_0)$ give estimates for the exponents
$\bar\delta$ and $\bar\Theta$. Within their error bars, the values for $p=0.1$ and 0.3
agree with the respective values for $p=0.2$ discussed in subsection
\ref{subsec:critical_behavior}.

Our simulations thus provide numerical evidence for the critical behavior of the
disordered 2D contact process being universal, i.e., being controlled by the same
infinite-randomness fixed point for all disorder strength. Note that the bare disorder can
be considered weak for $p=0.1$ as the difference of the critical infection rate from its
clean value is small, $(\lambda_c-\lambda_c^0)/\lambda_c^0 \approx 0.12$.

\subsection{Griffiths region}

\begin{figure}[tb]
\centerline{\includegraphics[width=8.cm]{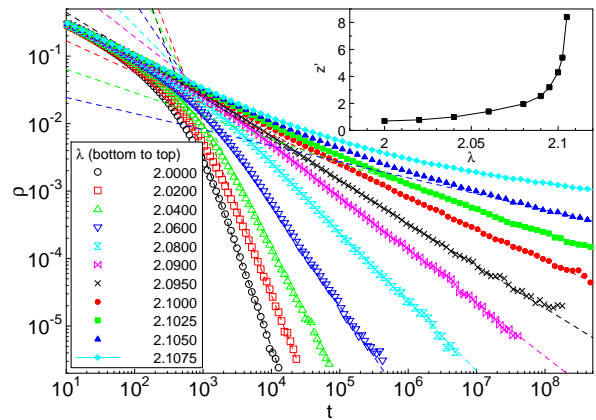}}
\caption{(Color online) $\ln \rho(t)$ versus $\ln t$ for several infection rates in the Griffiths
         region, $\lambda_c^0 < \lambda < \lambda_c$ ($p=0.2$, 13 disorder
         realizations with $L=8000$ for $\lambda>2.1$, 250 realizations with
         $L=2000$ for $\lambda \le 2.1$ ). The dashed lines are
         power-law fits of the long-time behavior to (\ref{eq:griffithspower}).
         Inset: Resulting dynamical exponent $z'$ as a function of $\lambda$. }
\label{fig:griffiths}
\end{figure}

In addition to the critical point, we have also studied the Griffiths region
$\lambda_c^0 < \lambda < \lambda_c$ in some detail. Figure \ref{fig:griffiths}
shows a double-logarithmic plot of the density $\rho(t)$ (starting from a
fully infected lattice) for several infection rates in this region.
For all these $\lambda$, the long-time behavior of the density follows the expected
power law over several order of magnitude in time and/or density.
The values of the (non-universal) dynamical exponent $z'$ can be determined
by fitting the long-time behavior to eq.\ (\ref{eq:griffithspower}). The inset of Fig.\ \ref{fig:griffiths}
shows that $z'$ diverges with $\lambda$ approaching the critical point $\lambda_c=2.1075$.
Fitting this divergence to the power law (\ref{eq:z_prime}) expected within the
infinite-randomness scenario, we find $\psi\nu_\perp \approx 0.61$ in good agreement
with the value extracted from the scaling analysis of the density in subsection
\ref{subsec:critical_behavior}. Performing the same analysis with the survival
probability data shown in Fig.\ \ref{fig:Psoverview} gives analogous results.

 \subsection{Spatial correlations}

\begin{figure}[tb]
\centerline{\includegraphics[width=8.cm]{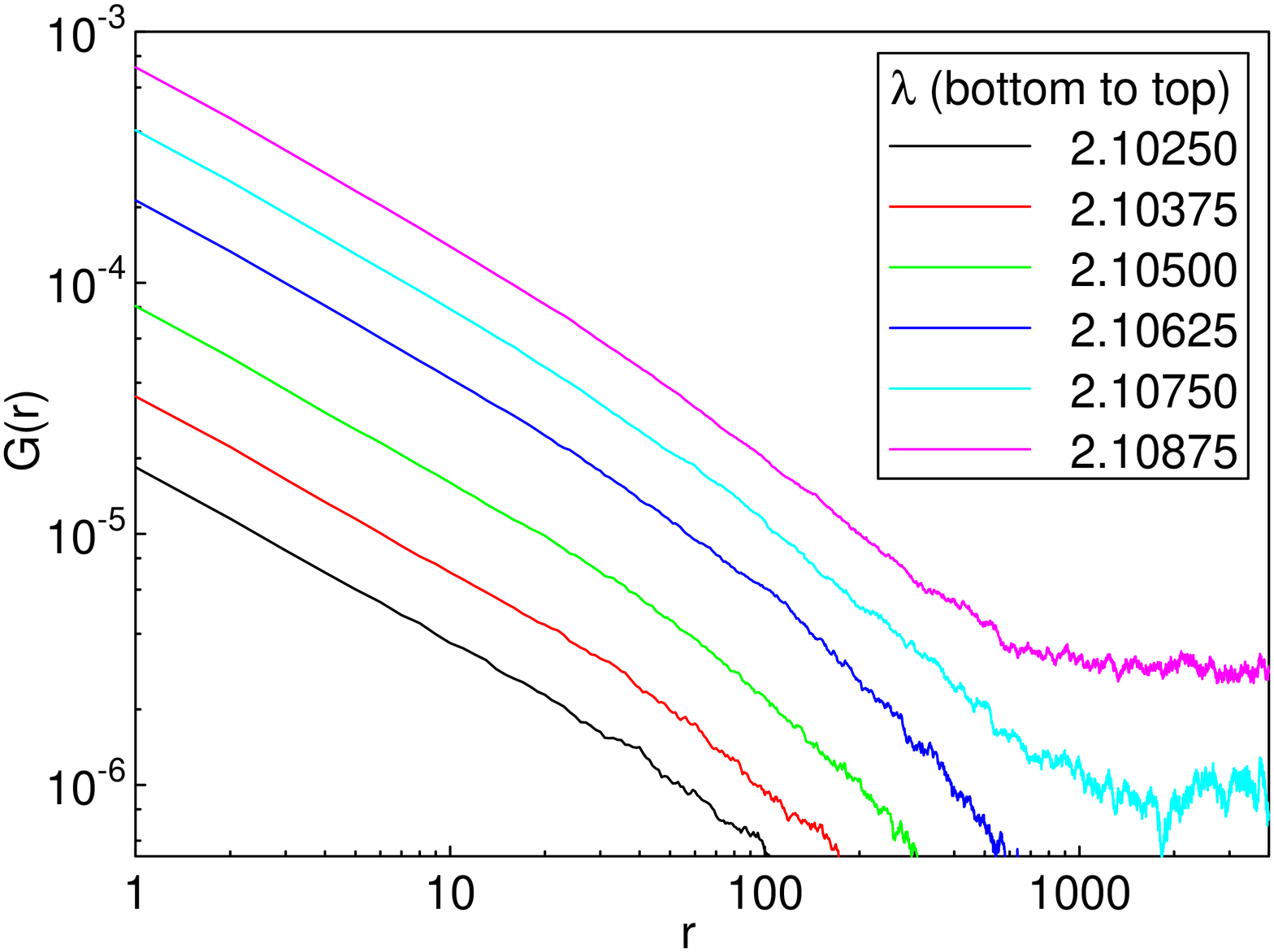}}
\caption{(Color online) $\ln G(r)$ vs. $\ln r$ for several infection rates
         $\lambda$ ($p=0.2$, 13 disorder realizations with $L=8000$).
         $G(r)$ is calculated from the configurations at late times
         of the time evolution ($t\ge 4\times 10^8$).}
\label{fig:correlations}
\end{figure}

To study the spatial correlations in the diluted contact process close to criticality,
we calculate the \emph{average} equal-time correlation function $G = \frac 1 N \sum_\mathbf{r} \langle n_\mathbf{r}(t)
n_0(t) \rangle$ from the late-time configurations arising in the
simulations starting from a fully infected lattice.
Scaling arguments analogous to those leading to (\ref{eq:C}) suggest the
scaling form
\begin{equation}
G(\Delta,\mathbf{r},\ln(t/t_0)) = b^{2\beta/\nu_\perp} G(\Delta b^{-1/\nu_\perp}, \mathbf{r}b,
\ln(t/t_0)b^\psi)~.
\label{eq:G}
\end{equation}
To find the true stationary behavior of $G$, one would ideally want to sample the
quasistationary state which is not directly available from our simulations for
$\lambda \le \lambda_c$. However, the finite-time data allow us to extract
the short-distance behavior up to a crossover distance $r_x(t)$ given by
the scaling combination $\ln(t/t_0)r_x^{-\psi} \gtrsim 1$.
Figure \ref{fig:correlations} shows the correlation function for infection rates
close to the critical point.
The data show a crossover at a length scale of about 200 compatible with the crossover
scale identified at the end of Section \ref{subsec:critical_point}
separating the pre-asymptotic behavior from the true asymptotic long-distance form. Unfortunately,
we cannot explore the asymptotic region because (i) fluctuations are too strong, and (ii)
for larger distances, we violate the condition $\ln(t/t_0)r_x^{-\psi} \gtrsim 1$
discussed above. Thus, the asymptotic functional form of the spatial correlation function
remains a task for the future.

\subsection{Nonequilibrium transition across the lattice percolation threshold}

So far, we have considered the generic transition occurring
for impurity concentrations
$p<p_c$. In the present subsection, we briefly discuss the nonequilibrium phase
transition of the contact process across the percolation threshold $p_c$
of the underlying lattice. It occurs at sufficiently large infection rates  and is
represented by the vertical line at $p_c$ in the phase diagram shown in
Fig.\ \ref{fig:pd}.

Vojta and Lee \cite{VojtaLee06}
developed a theory for this transition by combining the well-known critical behavior
of classical percolation with the properties of a supercritical contact process on a
finite-size cluster. They found that the behavior of the contact process on a diluted
lattice close to the percolation threshold follows the activated scaling scenario
of subsection \ref{subsec:activated}. However, the critical exponents differ from
those of the generic transition discussed above; they can be expressed as combinations
of the classical lattice percolation exponents $\beta_c$ and $\nu_c$ which are known
exactly in two space dimensions.
Specifically, the static exponents $\beta$ and $\nu_\perp$ are identical to their lattice
counterparts $\beta=\beta_c=5/36$ and $\nu_\perp=\nu_c=4/3$. The tunneling exponent $\psi$ is given
by $\psi=D_c=2-\beta_c/\nu_c=91/48$ where $D_c$ is the fractal dimension of the critical
percolation cluster (of the lattice). This implies an extremely small value for the exponent
controlling the critical density decay, $\bar\delta =\beta/(\nu_\perp\psi)=5/91$.

\begin{figure}[tb]
\centerline{\includegraphics[width=8.cm]{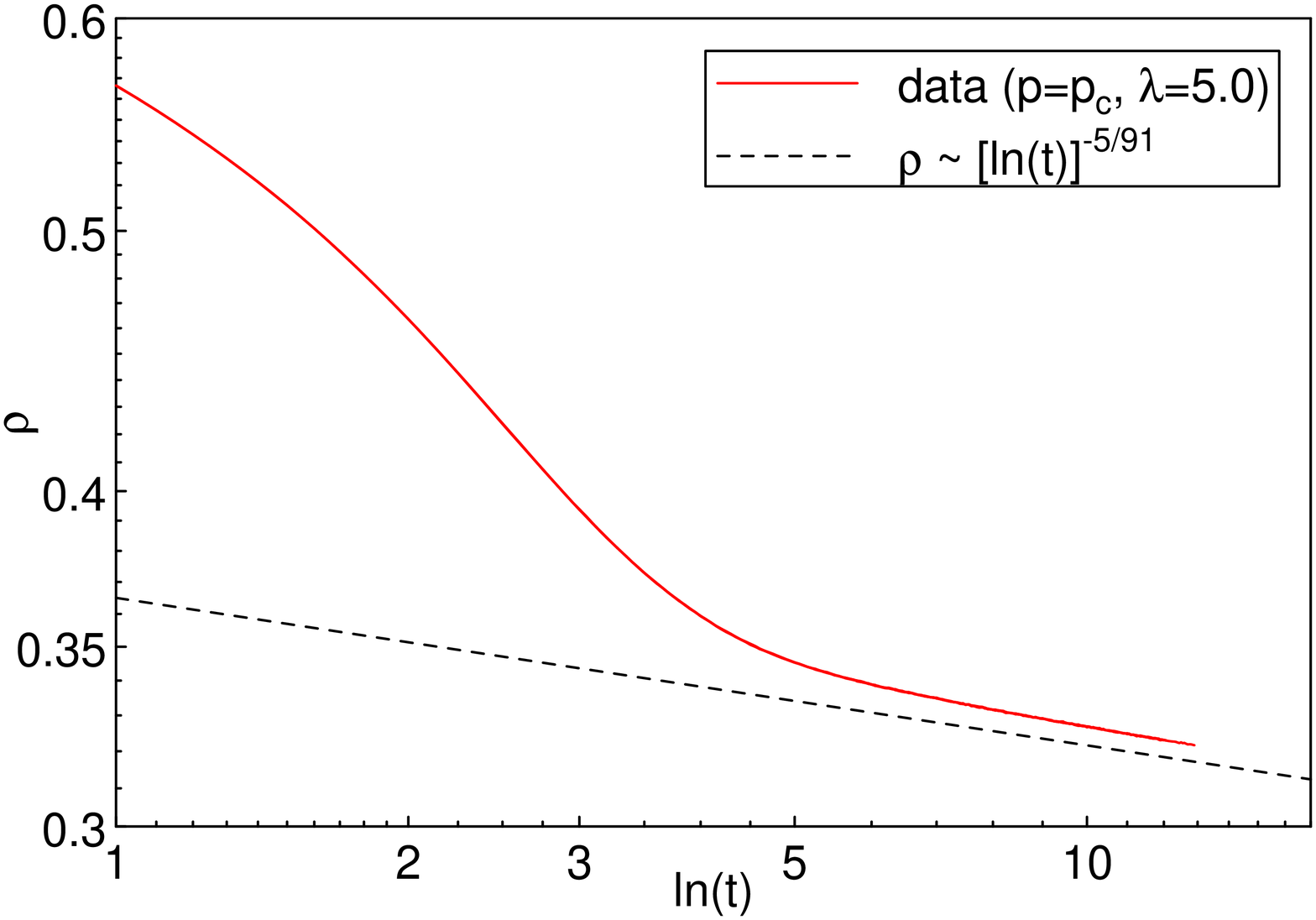}}
\caption{(Color online) $\ln \rho(t)$ versus $\ln \ln(t)$ for $p=p_c$, $\lambda=5.0$, and $L=8000$.
         We have simulated 3 samples, their differences are smaller than the line
         width. The dashed line represents the expected logarithmic long-time decay,
         $\rho \sim [\ln(t)]^{-\bar\delta}$ with arbitrary prefactor.  }
\label{fig:percolation}
\end{figure}

To test this prediction, we performed simulation runs at $p=p_c$ and $\lambda=5.0$,
starting from a fully infected lattice. The resulting time evolution of the density of
active sites (averaged over three samples of linear size $L=8000$) is presented in Figure
\ref{fig:percolation}.
The data show the behavior predicted in Ref.\ \cite{VojtaLee06}: a rapid initial
density decay towards a quasi-stationary density,
followed by a slow logarithmic time dependence due to the successive decays of
the contact process on larger and larger percolation clusters. The asymptotic behavior is
compatible with the predicted exponent value $\bar \delta =5/91$, however, determining an
accurate value of $\bar \delta$ from the simulation data is impossible because of the
extremely slow decay and the limited time range we can cover.

\section{Discussion and conclusions}
\label{sec:conclusions}

In this final section of the paper, we summarize our results and discuss some aspects of
our data analysis. We then briefly consider the diluted contact process in three space
dimensions. Finally, we relate our findings to a recent classification
of phase transitions with weak disorder.

\subsection{Summary}

We have performed large-scale Monte-Carlo simulations of the contact process on a
two-dimensional site-diluted lattice. We have determined the infection-dilution
phase diagram und studied both the generic nonequilibrium phase transition
for dilutions below the percolation threshold of the lattice and the transition
across the lattice percolation threshold.

Our main results is that the generic transition is controlled by an infinite-randomness
fixed point for all disorder strength investigated, giving rise to ultraslow activated
rather than power-law dynamical scaling. Based on two types of simulations, (i) spreading
of the infection from a single seed and (ii) simulations starting from a fully infected
lattice, we have determined the complete critical behavior. Our critical exponent values
are summarized in Table \ref{table:exponents}.
\begin{table}
\begin{tabular}{ccc}
\hline
critical exponent     &  value           & value\\
                       &  (generic)       & (percolation) \\
\hline
$\beta/\nu_\perp$            &  0.96(2)         & 5/48 \\
$\beta$                &  1.15(15)        & 5/36  \\
$\nu_\perp$                  &  1.20(15)        & 4/3  \\
$\psi$                 &  0.51(6)         & 91/48 \\
$\bar\delta$           &  1.9(2)          & 5/91 \\
$\bar\Theta$           &  0.15(3)         &       \\
\hline
\end{tabular}
\caption{Critical exponents for the nonequilibrium phase transitions
of the disordered two-dimensional contact process. The values for the
generic transition are from this work, the values for the percolation
transition are from Ref.\ \cite{VojtaLee06}. The numbers in brackets
give an error estimate of the last given digits.}
\label{table:exponents}
\end{table}
Note that the spatial correlation length exponent fulfills the Harris criterion
 \cite{Harris74}
inequality $d\nu_\perp\ge2$, as expected in a disordered system \cite{CCFS86}.

We have been able to obtain a rather accurate estimate for the finite-size scaling
exponent $\beta/\nu_\perp$. However, the other exponents are somewhat less accurate
despite our extensive numerical effort (system sizes up to 8000$\times$8000 sites and
times up to $10^{10}$). This is caused by three important characteristics of our
infinite-randomness critical point:

(i) The logarithmic time-dependencies associated
with activated scaling are not scale-free, they contain a microscopic time scale
$t_0$ which acts as a strong correction to scaling and must be estimated independently
(otherwise the exponent values will be seriously compromised).
Since our simulations cover only a limited range of $\ln(t)$, our values for $t_0$
are not very precise which limits the accuracy of the exponents $\bar\delta$,
$\bar\Theta$, and $\psi$.

(ii) The second problem consists in the extreme narrowness in $\lambda$ of the asymptotic
critical region for the weak to moderate disorder strengths we have considered. We
estimated a relative width of $|\lambda-\lambda_c|/\lambda_c \lessapprox 10^{-3}$ for
dilution $p=0.2$. This hampers the analysis of off-critical data and thus limits the
accuracy of the exponents $\beta$ and $\nu_\perp$

(iii) The crossover of the time evolution to the asymptotic critical behavior occurs
very late. For $p=0.2$ the crossover
time $t_x$ is about $10^4$. Moreover, the diameter of the active cloud spreading from a single
seed has reached about 200 at that time. Thus, successful simulations require both large
systems and long times.

Let us finally comment on the universality of the critical exponents. We have studied
three different dilution values, $p=0.1$, 0.2, and 0.3, ranging from weak to moderate
disorder as judged by the shift in the critical infection rate with respect to its
clean value. While the crossover time
$t_x$ and the microscopic time scale $t_0$ vary significantly with dilution, the asymptotic
critical exponents remain constant within their error bars (including the rather
accurate finite-size scaling exponent $\beta/\nu_\perp$). Our data thus provide no indication
of nonuniversal continuously varying exponents. However, it must be emphasized that due to
the limited precision, some variations cannot be rigorously excluded.

In the case of the transition of the contact process across the lattice percolation
threshold, our simulation
data support the theory developed in Ref.\ \cite{VojtaLee06}: The dynamical critical
behavior is activated, with the critical exponents given by combinations of the exactly
known lattice percolation exponents.

\subsection{Three dimensions}

While the existence of infinite-randomness critical points in one and two spatial
dimensions is well established in the literature, the situation in three dimensions is
less clear. Within a numerical implementation of the strong-disorder renormalization
group, Montrunich et al. \cite{MMHF00} found that the flow is towards strong disorder.
However, the data were not sufficient for analyzing the critical behavior quantitatively.

\begin{figure}[tb]
\centerline{\includegraphics[width=8.cm]{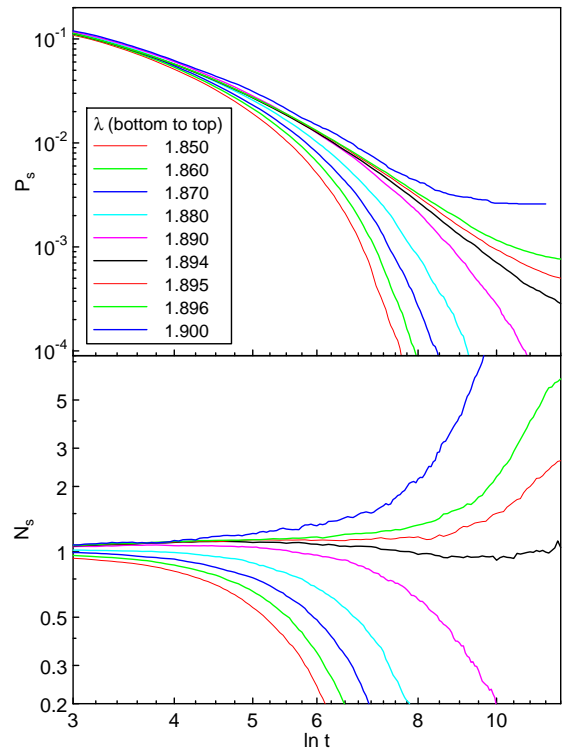}}
\caption{(Color online) $\ln P_s(t)$ and $\ln N_s(t)$ versus $\ln \ln t$ for the three-dimensional
         contact process on a diluted lattice ($p=0.3$, $L=400$, at least 5000 disorder
         realizations with 128 trials each).}
\label{fig:3d}
\end{figure}

We are currently performing Monte-Carlo simulations of the three-dimensional contact
process on a diluted lattice analogous to those reported in Sec.\ \ref{sec:mc}.
Preliminary results shown in Fig.\ \ref{fig:3d} suggest that the phase transition
scenario is very similar to that in one and two-dimensions: the transition appears to
be controlled by an infinite-randomness fixed point with activated dynamical scaling.
A detailed quantitative analysis of the critical behavior requires significantly longer
times and larger systems. It will be published elsewhere.

\subsection{Conclusions}

Recently, a general classification of phase transitions in quenched disordered systems
with short-range interactions has been suggested \cite{VojtaSchmalian05,Vojta06}. It is
based on the effective dimensionality $d_{\rm eff}$ of the defects (or, equivalently,
the rare regions.) Three cases can be distinguished.

(A) If $d_{\rm eff}$ is below the lower critical dimension $d_c^-$ of the problem, the
rare region effects are exponentially weak, and the critical point is of
conventional type.
(B) In the second class, with $d_{\rm eff}=d_c^-$, the Griffiths effects are of
strong power-law type and the critical behavior is controlled by an infinite-randomness
fixed point with activated scaling.
(C) For  $d_{\rm eff}>d_c^-$, the rare regions can undergo the phase
transition independently from the bulk system. This leads to a destruction of the sharp
phase transition by smearing \cite{Vojta03a}.

The results of this paper agree with this general classification scheme as $d_{\rm eff}=d_c^-=0$
(this corresponds to rare regions being marginal with their life time depending
exponentially on their size) leading to class B. In contrast, the contact process with
\emph{extended} (line or plane) defects falls into class C \cite{Vojta04,DickisonVojta05}.

We conclude by pointing out that the unconventional behavior found in this paper may
explain the striking absence of directed percolation scaling \cite{Hinrichsen00b} in at
least some of the experiments. However, the extremely slow dynamics and narrow critical region
will prove to be a challenge for the verification of the activated scaling scenario not
just in simulations but also in experiments.
We also emphasize that our results are of importance beyond absorbing state transitions.
Since the strong-disorder renormalization group predicts our transition to be in the
universality class of the two-dimensional random transverse-field Ising model, the critical
behavior found here should be valid for this entire universality class.

\section*{Acknowledgements}

This work has been supported in part by the NSF under grant no. DMR-0339147, by Research
Corporation, and by the University of Missouri Research Board. We gratefully acknowledge
discussions with R. Dickman, J. Hoyos, and U. T\"{a}uber as well the hospitality of the
Aspen Center for Physics during part of this research.

\bibliographystyle{apsrev}
\bibliography{../00Bibtex/rareregions}

\begin{thebibliography}{54}
\bibitem[{Bib({\natexlab{b}})}]{Bibnote2}
\bibinfo{note}{We define $p$ is the fraction of sites removed rather than the
  fraction of sites present.}
\bibitem[{Bib({\natexlab{c}})}]{Bibnote3}
\bibinfo{note}{At more general absorbing state transitions, e.g., with several
  absorbing states, the survival probability scales with an exponent $\beta '$
  which may be different from $\beta $ (see, e.g., \cite {Hinrichsen00}).}
\bibitem[{Bib({\natexlab{d}})}]{Bibnote4}
\bibinfo{note}{This relation relies on hyperscaling; it is only valid below the
  upper critical dimension $d_c^+$, which is four for directed percolation}.
\end{thebibliography}
\end{document}